\documentclass[twocolumn,showpacs,preprintnumbers,amsmath,amssymb,floats,citeautoscript]{revtex4-1}
\usepackage{graphicx}
\usepackage{dcolumn}
\usepackage{bm}
\usepackage{txfonts}
\usepackage{multirow}
\usepackage{color}
\usepackage{here}

\begin{document}
 
\title{Nonlinear transport associated with spin-density-wave dynamics in Ca$_3$Co$_{4}$O$_9$}

\author{N.~Murashige}
\author{F.~Takei}
\author{K.~Saito}
\author{R.~Okazaki}

\affiliation{Department of Physics, Faculty of Science and Technology, Tokyo University of Science, Noda 278-8510, Japan}

\begin{abstract}

We have carried out the transient nonlinear transport measurements on the 
layered cobalt oxide Ca$_3$Co$_{4}$O$_9$, in which a 
spin density wave (SDW) transition is proposed at $T_{\rm SDW} \simeq 30$~K.
We find that, below $T_{\rm SDW}$,
the electrical conductivity systematically varies with both the applied current and the  time,
indicating a close relationship between the observed nonlinear conduction and the 
SDW order in this material.
The  time dependence of the conductivity is  well analyzed 
by considering the dynamics of SDW which involves a low-field deformation and a sliding motion above a threshold field.
We also measure the transport properties of the isovalent Sr-substituted systems
to examine an impurity effect on the nonlinear response,
and discuss the obtained threshold fields in terms of thermal fluctuations of the SDW order parameter.

\end{abstract}

\maketitle

\section{introduction}

The quasi-two-dimensional cobalt oxide Ca$_3$Co$_{4}$O$_9$ provides a fascinating research playground  
to examine the unusual transport of 
strongly correlated electrons,
which is characterized by the intimate coupling with
its complicating structural and magnetic properties.
This compound
consists of two subsystems of
CdI$_2$-type CoO$_2$ and rocksalt-type Ca$_2$CoO$_3$ layers,
alternately stacked
along the $c$ axis \cite{Masset2000,Miyazaki2002}.
The former is responsible for the charge transport
and 
the latter behaves as the
charge reservoir to supply holes into the conduction layer \cite{Yang2008,Tanabe2016,Mizokawa2005,Klie2012}.
As is
distinct from other layered systems such as the high-$T_c$ cuprates,
these subsystems
have different $b$ axis parameters ($b_1$ being for the rocksalt and $b_2$ for the CoO$_2$ layers).
The misfit ratio is $b_1/b_2\simeq1.62$, thus
[Ca$_2$CoO$_3$][CoO$_2$]$_{1.62}$ is more realistic,
whereas 
we refer to this system as 
the approximate formula 
Ca$_3$Co$_{4}$O$_9$. 

Ca$_3$Co$_{4}$O$_9$
exhibits
several magnetic transitions with respect to temperature.
Around room temperature, 
this material is paramagnetic 
and the cobalt ions take the low spin state.
Above room temperature, the magnetic susceptibility shows an
anomaly at $T=380$~K,
although the spin-state nature at higher 
temperature is not elucidated \cite{Sugiyama2002,Sugiyama2003,Wakisaka2008}.
With decreasing temperature, 
a short-range order of spin density wave (SDW) develops below 100~K, and 
the long-range order appears below $T_{\rm SDW} \simeq 30$~K \cite{Sugiyama2002,Sugiyama2003}. 
The SDW coexists with 
a ferrimagnetism below $T_{\rm FR}=19$~K.

The unsolved issue is  how such complicating magnetism relates to 
the transport properties.
This compound shows
a metallic resistivity  with
a large Seebeck coefficient reaching 
$S\simeq 130$~$\mu$V/K
at around room temperature \cite{Masset2000},
which is qualitatively understood with
a large spin entropy 
flow associated with the charge (hole) hopping
among the low-spin cobalt ions \cite{Koshibae2000,Koshibae2001,Hejt2015}.
With lowering temperature,
the resistivity shows a Fermi-liquid behavior with $\rho(T)=\rho_0+AT^2$ 
around 120~K (fig.1) as is widely seen in strongly correlated electron systems,
but 
below $T\sim70$~K, it increases
like an insulator while the Seebeck coefficient shows a metallic behavior \cite{Limelette2005}.
Such a resistivity upturn may invoke 
the variable range hopping transport 
with $\rho(T)\propto\exp(T^{-\alpha})$ $(0<\alpha<1)$ \cite{Bhaskar2014},
but 
it is also claimed that 
the measured temperature dependence of the resistivity is much milder 
and that the fitting range is narrow. 
Limelette {\it et al.} have shown a large negative magnetoresistance at low temperatures,
indicating a spin-dependent scattering mechanism \cite{Limelette2008}.
Hsieh {\it et al.} have found that the conductivity is well scaled with the Seebeck coefficient
for a two-dimensional metal, 
suggesting a reduction of the carrier density with decreasing temperature 
due to a pseudogap opening 
associated with the SDW formation \cite{Hsieh2014}.
Such a close relation between the resistivity upturn and the SDW order
is also seen in the related layered oxide Na$_x$CoO$_2$ \cite{Sugiyama2004,Lee2006}.
On the other hand, recent theoretical study suggests that 
such a resistivity increase is not necessarily related to the magnetic transitions \cite{Lemal2017},
remaining the underlying origin of the transport properties unclear.

\begin{figure}[b]
\includegraphics[width=8cm]{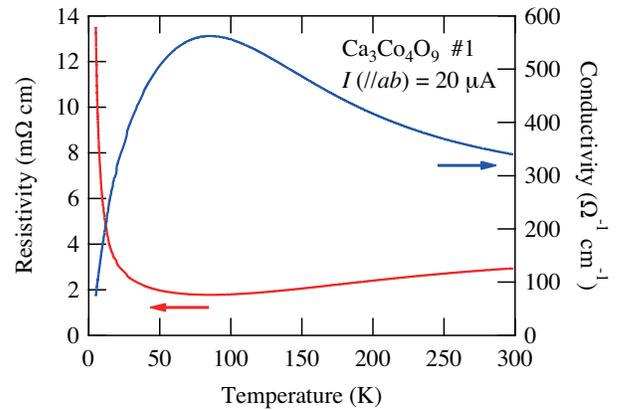}
\caption{(Color online).
Temperature variations of the in-plane resistivity (left axis) and the conductivity (right axis)
of Ca$_3$Co$_{4}$O$_9$ single crystal (sample \#1 with low contact resistance) measured with a low excitation current of $I=20$~$\mu$A.
}
\end{figure}

\begin{figure*}[t]
\includegraphics[width=17cm]{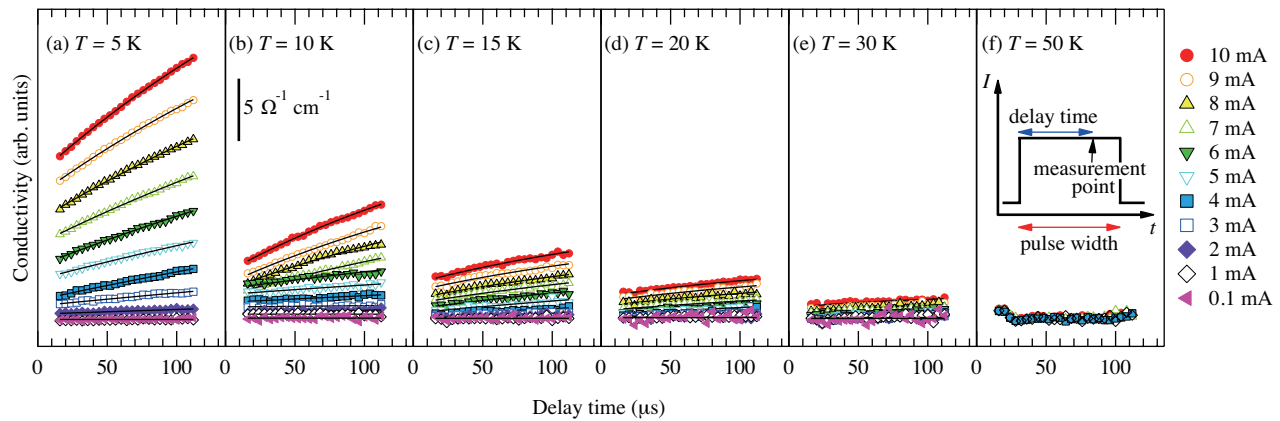}
\caption{(Color online).
(a-g) Time variations of the conductivity of Ca$_3$Co$_{4}$O$_9$ single crystal (sample \#1 with low contact resistance)
measured at several temperatures.
The scale bar shown in (b) is common to all the data in (a-f), and the data are not shifted for each panel.
The solid curves show the fitting results. 
The inset  depicts a schematic curve of the applied  current pulse. 
}
\end{figure*}

Here,
we have performed the nonlinear conduction measurements on  Ca$_3$Co$_{4}$O$_9$
single crystals to examine the electric-field response of the SDW.
We measure the time dependence of the conductivity under a short current pulse 
at several temperatures.
A clear nonlinear conduction, 
characterized by 
a systematic change in the conductivity by applied current and time,
has been observed 
below $T_{\rm SDW}$,
and 
the time dependence of the conductivity is well understood
by considering the dynamics of SDW including a low-field deformation and a high-field sliding motion.
This result indicates an intimate coupling of the low-temperature transport with the 
SDW order in this system.
Moreover, to reveal the impurity effect, the nonlinear conduction of the isovalent Sr-substituted 
Ca$_{3-x}$Sr$_x$Co$_{4}$O$_9$ is also measured.
We discuss the thermal fluctuation  effect on the SDW order to explain 
the obtained temperature and Sr-content dependences of the threshold field.

\section{experiments}

The experiments were performed using 
Ca$_{3-x}$Sr$_x$Co$_{4}$O$_9$ single crystals ($x=0,0.1,0.2$) with typical sample dimensions of $1\times1\times0.04$\,mm$^3$ 
grown by a flux method \cite{Ikeda2016}. 
Powders of CaCO$_3$ (99.9\%), SrCO$_3$ (99.9\%), and Co$_3$O$_4$ (99.9\%)
were mixed in a stoichiometric ratio and
calcined two times in air at 1173~K for 24~h with intermediate grindings.
Then KCl  (99.999\%) and K$_2$CO$_3$  (99.999\%) powders mixed with a molar ratio of $4:1$ was added 
with the calcined powder as a flux. 
The concentration of Ca$_{3-x}$Sr$_x$Co$_{4}$O$_9$ was set to be 1.5\% in molar ratio. 
The mixture was put in an alumina crucible and 
heated up to 1123~K in air with a heating rate of 200~K/h.
After keeping 1123~K  for 1~h, 
it was slowly cooled down  with a rate of 1~K/h, 
and at 1023~K, the power of the furnace was switched off.
As-grown samples were rinsed in distilled water to remove the flux.

The resistivity was measured using a standard four-probe method
along the in-plane direction.
The transient measurement was performed using a Keithley 6221 current source and a 2182A nanovoltmeter, 
which were synchronously operated in a built-in pulse mode.
The pulse width of the excitation current was set to be 160~$\mu$s.
The pulse interval was set to be 20~ms, which is sufficiently long to reduce the Joule heating.
The time dependence of the sample voltage
was measured 
by changing the delay time
from the leading edge of the current pulse  for the voltage measurement,
as is schematically depicted in the inset of Fig. 2(f).
Owing to the limitation of the equipment, the minimum delay time is about 20 $\mu$s.

The electrodes were made by two kinds of silver paints
to check the effect of the contact resistance.
Except for Ca$_3$Co$_{4}$O$_9$ sample \#2, 
we used Dupont 6838, which was cured above 473~K for 2~h,
and achieved a relatively low contact resistance of $R_{\rm contact}\sim 10$~$\Omega$
at low temperature of $T=5$~K.
On the other hand, for Ca$_3$Co$_{4}$O$_9$ sample \#2, 
we used Dupont 4922 cured at room temperature.
Compared to the former case, 
the contact resistance was  high ($R_{\rm contact}\sim 100$~$\Omega$ at $T=5$~K).
The effect of the contact resistance on the nonlinear conduction will be discussed later.

\section{results and discussion}

Figure 1 shows the temperature dependences of the resistivity $\rho$ and 
the conductivity $\sigma = \rho^{-1}$ of 
Ca$_3$Co$_{4}$O$_9$ single crystal (sample \#1)
measured with a low excitation current of $I=20$~$\mu$A.
As is seen in the previous studies \cite{Masset2000}, 
the resistivity shows 
a metallic temperature variation above 100~K but 
 displays an insulating behavior at low temperatures.
Figures 2(a-f) depict the time variations of the conductivity of 
Ca$_3$Co$_{4}$O$_9$ single crystal (sample \#1) measured at 
several temperatures 
in the insulating regime.
At $T=5$~K [fig. 2(a)], 
the conductivity increases with increasing time and 
applied current,
indicating a presence of nonlinear conduction in this system.
The nonlinearity becomes small rapidly
with increasing temperature, 
and it completely disappears at $T=50$~K 
as is seen in fig. 2(f).

Here, 
we first note that the observed change in the conductivity 
does not stem from the Joule heating under the large current,
which often appears
in the nonlinear conduction measurements \cite{Okazaki2013}.
To check this extrinsic effect,
we estimate the maximum temperature increase $\Delta T_{\rm max}$
by assuming that all the thermal energy produced by the Joule heating is consumed 
for the increase of the sample temperature.
In the present study,
the sample resistance ($<1$ $\Omega$) is much smaller than 
the contact resistance $R_{\rm contact}\sim10$~$\Omega$,
hence the contact resistance determines the Joule heating power.
We then calculate $\Delta T_{\rm max}$ as
$\Delta T_{\rm max} = R_{\rm contact}I^2\Delta t/CV$,
where $I$ is the applied current, 
$\Delta t$ the pulse width,
$C$ the heat capacity at $T=5$~K \cite{Limelette2005},
and 
$V$ the sample volume,
and obtain
$\Delta T_{\rm max} \sim 0.1$~K for $I = 10$~mA and $T=5$~K.
Since $d\sigma/dT\sim20$~$\Omega^{-1}$cm$^{-1}$K$^{-1}$
around $T=5$~K, as is seen in  
fig. 1, $\Delta T_{\rm max} \sim 0.1$~K  corresponds to 
the conductivity change of 
2~$\Omega^{-1}$cm$^{-1}$.
This maximum estimation is 
much smaller than the observed conductivity change as shown in 
fig. 2(a),
indicating an intrinsic nonlinearity in this system.
We also examine 
the Joule heating effect by using the sample \#2 with high contact resistance,
which will be discussed later.

\begin{figure}[t]
\includegraphics[width=7.5cm]{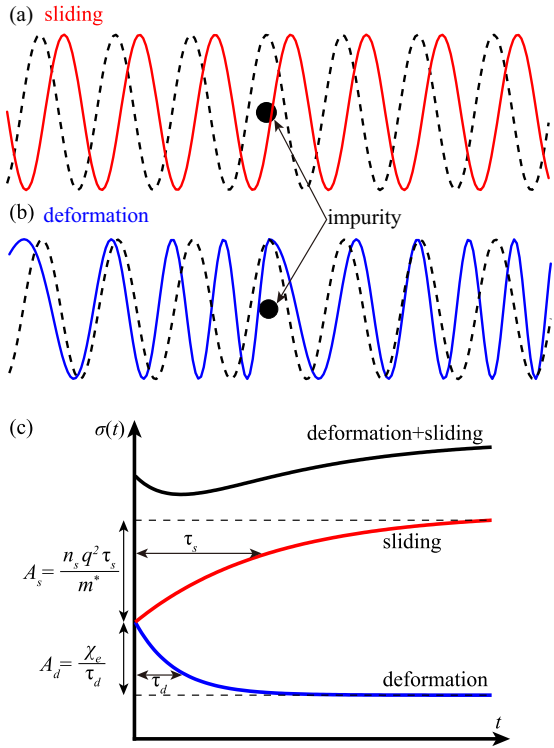}
\caption{(Color online).
Schematic pictures of (a) sliding motion and (b) deformation of spin density wave (SDW)
around an impurity expressed as the solid circle.
The dashed and solid curves show the SDW without and with electric field, respectively.
(c) Schematic time variations of the conductivity.
The red and blue curves are conductivities from sliding motion and deformation of SDW, respectively,
and the black solid curve represents the total conductivity.
}
\end{figure}

Since the nonlinear conduction becomes negligibly small above the 
SDW transition temperature $T_{\rm SDW}\sim$30~K,
we then analyze 
the present data by considering the dynamics of density wave.
In contrast to a charge density wave (CDW),
the charge density is spatially constant in a SDW.
However, 
a SDW can be regarded as 
a composition of two out-of-phase CDW's formed by
spin-up and spin-down electrons.
Near an impurity site, the SDW will be pinned by inducing a 
distortion of the total electron density because 
the up- and down-spin components of the charge density 
deform differently.
This is basically reduced to the coupling of the impurities to
the second-order harmonic CDW that coexists with the SDW \cite{Tua1985,Tomic1989}.
A similar electric field response to that of CDW is therefore
expected in a SDW system 
and has experimentally observed \cite{Monceau}.

Now 
the classical equation of motion of a density wave under electric field $E$ is expressed as
\begin{equation}
\frac{d^2x}{dt^2}=-\frac{1}{\tau_s}\frac{dx}{dt}-\frac{\omega_0^2}{2k_F}\sin2k_Fx+\frac{qE}{m^*},
\end{equation}
where $x$ is the displacement, 
$\tau_s$ a relaxation time,
$\omega_0$ a pinning frequency,
$k_F$ a Fermi wavenumber,
$q$ a charge, 
and 
$m^*$ an effective mass for the density wave \cite{Gruner,Monceau}.
Under strong electric field, the density wave exhibits a 
collective sliding motion as is schematically depicted in fig. 3(a).
This motion carries an additional electrical current 
and increases
the conductivity.
In this case, the electric potential is much larger than the pinning potential.
Then the restoring term may be neglected, and the equation is reduced to
\begin{equation}
\frac{dv}{dt}=-\frac{v}{\tau_s}+\frac{qE}{m^*},
\end{equation}
where $v = dx/dt$ \cite{Sasaki1996}.
This viscous-resistance-type equation is simply solved as
\begin{equation}
v(t) = \frac{q\tau_s E}{m^*} \left[1-\exp\left(-\frac{t}{\tau_s}\right)\right].
\end{equation}
Thus the conductivity carried by the sliding motion of density wave 
$\sigma_s$ is given as
\begin{equation}
\sigma_s(t) = 
\frac{n_sqv}{E} =
A_s\left[1-\exp\left(-\frac{t}{\tau_s}\right)\right],
\end{equation}
where $A_s=n_s{q}^2\tau_s/m^*$ and 
$n_s$ is a carrier density participating the sliding motion of density wave.
The time dependence of the conductivity $\sigma_s(t)$ is schematically 
shown by the red curve in fig. 3(c).
The total conductivity is then given as $\sigma_s+\sigma_0$,
where $\sigma_0$ is an offset term, and 
the fitting well reproduces the experimental results as
shown by the solid curves in figs. 2(a-f).
The obtained fitting parameters will be discussed later.

\begin{figure}[t]
\includegraphics[width=8.5cm]{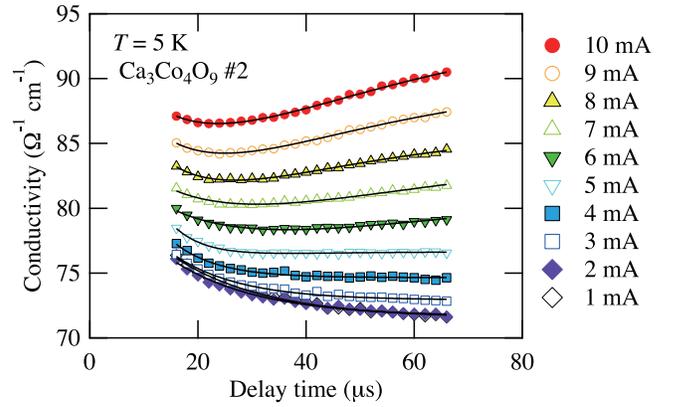}
\caption{(Color online).
Time variations of the conductivity of Ca$_3$Co$_{4}$O$_9$ single crystal (sample \#2
with high contact resistance)
measured at $T=5$~K.
The solid curves show the fitting results. For details see text.
}
\end{figure}

We also examine the nonlinear conduction in 
Ca$_3$Co$_{4}$O$_9$ sample \#2,
in which the contact resistance is 10 times larger than 
that in sample \#1.
Figure 4
shows the time variations of the conductivity of Ca$_3$Co$_{4}$O$_9$ sample \#2
measured at $T=5$~K.
In contrast to the results of sample \#1 shown in figs. 2(a-f),
the conductivity first decreases with increasing time 
in the short time range below about 30 $\mu$s.
This behavior is opposite to the heating effect since the conductivity should increase with heating at this temperature range.
It then turns to increase with time
at the high current range like the sliding motion 
as is discussed before.
To examine the heating effect in this range,
we compare the conductivity change by currents.
At the delay time of 60 $\mu$s, 
the conductivity changes between 1~mA and 10~mA are 
17~$\Omega^{-1}$cm$^{-1}$ for sample \#1 (fig. 2(a)) and 
18~$\Omega^{-1}$cm$^{-1}$ for sample \#2 (fig. 4), 
which are almost the same. 
If the heating effect is dominant, the conductivity in the sample \#2 should largely change due to the 
10-times difference in the contact resistance.
This result again indicates a negligible heating effect in the present study.

To explain the result of sample \#2, we now consider the field-induced deformation of SDW 
in addition to the sliding motion.
As is schematically depicted in fig. 3(b),
at low applied fields, a density wave is pinned by impurities and is deformed \cite{Sasaki1996}.
Here the resulting spatial charge modulation produces
a time-dependent electric polarization $P(t)$.
In density-wave materials, the Debye description is a first approximation 
for the dielectric relaxation process \cite{Cava1984,Mihaly1991},  
hence we use
\begin{equation}
\frac{dP(t)}{dt}=-\frac{P(t)}{\tau_d}+\frac{\chi_eE}{\tau_d},
\end{equation}
where $\tau_d$ is a relaxation time for the deformation process
and $\chi_e$ is an electric susceptibility. 
This is solved as 
$P(t) = \chi_eE \left[1-\exp\left(-t/\tau_d\right)\right]$.
We thus obtain the polarization current $j_P=dP/dt$ and 
the deformation-induced conductivity $\sigma_d =j_P/E$
expressed as
\begin{equation}
\sigma_d(t) = \frac{\chi_e}{\tau}\exp\left(-\frac{t}{\tau}\right),
\end{equation}
as is schematically drawn by the blue curve in fig. 3(c).

\begin{figure}[t]
\includegraphics[width=8cm]{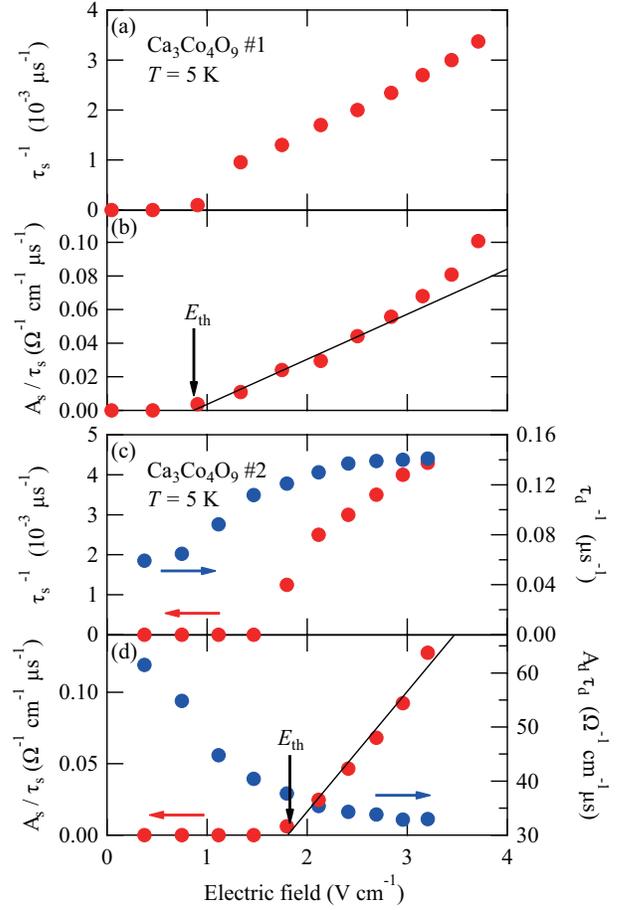}
\caption{(Color online).
Electric field dependence of (a) the scattering rate for the sliding motion $\tau_s^{-1}$
and (b) $A_s/\tau_s$ $(\propto n_s)$ for the sample \#1 at $T=5$~K.
The solid arrow represents the threshold field $E_{th}$.
(c) Electric field dependence of the scattering rate for the sliding motion (red circles, left axis) and the deformation process 
(blue circles, right axis)
for the sample \#2 measured at $T=5$~K.
(d) Electric field dependence of $A_s/\tau_s$ $(\propto n_s)$ (red circles, left axis) and the electric susceptibility 
$A_d\tau_d$ $(=\chi_e)$
(blue circles, right axis)
for the sample \#2 measured at $T=5$~K.
}
\end{figure}

Therefore the total conductivity is given as 
$\sigma(t) =\sigma_s(t) + \sigma_d(t)+\sigma_0$.
The time dependence of the total conductivity is schematically displayed 
by black curve in fig. 3(c),
and the fitting results for the sample \#2 are shown by solid curves in fig. 4, 
which also well reproduce the experimental data.
Note that the deformation-induced conductivity has not been monitored in the sample \#1,
as is seen in figs. 2(a-f).
We consider that, since the polarization current is induced by the electric dipoles pinned
at the interface between the sample and the electrode, 
it will be sensitive to the amount of the impurities at the interface,
hence the deformation-induced conductivity has been observed only 
in the sample \#2 with high contact resistance.

In figs. 5(a-d),
we summarize the fitting parameters obtained  at $T=5$~K.
Figures 5(a) and 5(c) display the
electric field dependences of the 
scattering rate $\tau_s^{-1}$ for the sample \#1 and 
the sample \#2 (left axis), respectively.
Here the electric field is obtained by averaging at each applied current.
For both samples,
the scattering rate becomes large with increasing  electric field
since the velocity of the sliding motion increases in the high fields.
Figures 5(b) and 5(d) show 
$A_s/\tau_s$
as a function of the
electric field for the sample \#1 and 
the sample \#2 (left axis), respectively.
Since 
$A_s=n_s{q}^2\tau_s/m^*$, 
$A_s/\tau_s$ is proportional to the carrier density which participates the sliding motion of SDW.
At low fields, $A_s/\tau_s$ 
is almost zero due to the pinning of SDW by impurities.
Above the threshold field $E_{th}$,
$A_s/\tau_s$ increases 
with increasing  electric field
since the number of depinned SDW increases 
in the high-field range.
We also stress that 
the electric field dependences of $\tau_s$ and $A_s/\tau_s$ do not differ significantly
between the samples \#1 and \#2,
because the conductivity changes are almost the same as discussed before.

We plot the electric field variations of 
$\tau_d^{-1}$ and $A_d\tau_d$ obtained 
for sample \#2 at $T=5$~K
in the right axes of figs. 5(c) and 5(d),
respectively.
The relaxation time for the deformation decreases with increasing electric field
as is the case in the sliding motion.
The electric susceptibility $A_d\tau_d$ $(=\chi_e)$
gradually decreases with increasing electric field and shows a weak electric field dependence 
above the threshold field.
This indicates that 
the polarization becomes small with increasing electric field owing to the
depinning of the SDW.
The background signal of $A_d\tau_d$ above the threshold field may originate from 
an extrinsic origin such as an unwanted capacitance component 
at the electrode due to the high contact resistance of the sample \#2.

\begin{figure}[t]
\includegraphics[width=8.5cm]{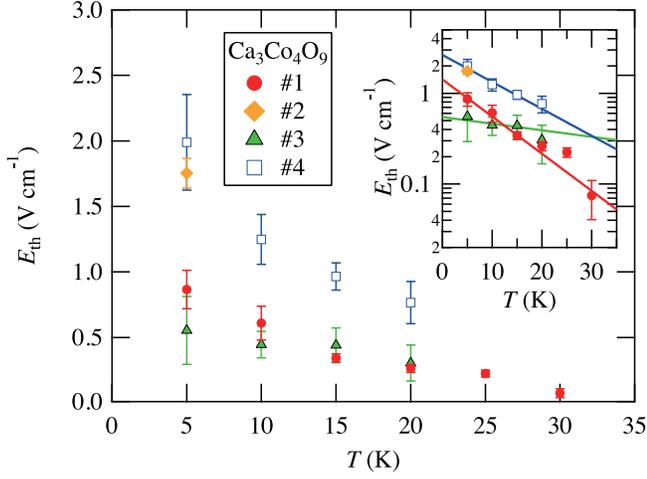}
\caption{(Color online).
Temperature dependence of the threshold field $E_{th}$ of
four samples of Ca$_3$Co$_{4}$O$_9$.
The inset shows $E_{th}$ vs $T$
in log-linear scale.
The solid lines represent the fitting results using 
$\ln E_{th}(T)=\ln E_{th}(0)-T/T_0$.
}
\end{figure}

The temperature dependence of the threshold field $E_{th}$ measured for
several samples is plotted in Fig. 6. 
The threshold field decreases with increasing temperature,
indicating a thermal fluctuation effect on the SDW order.
The threshold field under thermal fluctuations is theoretically given as
\begin{equation}
E_{th}(T)/E_{th}(0)=\exp(-T/T_0)
\end{equation}
for a low temperature range of $T\lesssim \frac{1}{2}T_{\rm SDW}$,
where 
$T_0 \sim \eta^2(k_F\xi)T_F$, 
$\eta=v_F^c/v_F^{ab}<1$ is the anisotropy of the Fermi velocity,
$\xi$ the coherence length of the SDW,
$T_F$ the Fermi temperature \cite{Maki1986}.
We then plot $E_{th}$ vs $T$ data in log-linear scale 
in the inset of fig. 6.
The data are fitted for  
a low-temperature range $T\leq 15$~K,
and as shown by the fitting lines,
$E_{th}(T)$ well obey this behavior,
suggesting the dominant thermal fluctuation effect.

Although $E_{th}(T)$  in the parent compounds
shows qualitatively similar temperature variations,
the resultant fitting parameters, $E_{th}(0)$ and $T_0$,
are strongly sample-dependent as is obviously seen in fig. 6, 
indicating that the uncontrollable impurities such as the oxygen defects may affect 
$E_{th}(T)$.
Thus we have measured 
the nonlinear conduction in
the isovalent compounds
Ca$_{3-x}$Sr$_x$Co$_{4}$O$_9$ ($x=0.1,0.2$)
to clarify the impurity effect on the 
threshold field.
Similar nonlinear conduction has also been observed 
in the Sr-substituted systems, and 
the obtained threshold fields are presented
in fig. 7(a) as a function of temperature.
Since the $\mu$SR study has revealed that 
the Sr substitution does not affect 
the SDW transition temperature \cite{Sugiyama2002},
we fit the data using eq. (7) for the temperature range 
of $T\leq 15$~K, 
as is depicted by the solid lines.

\begin{figure}[t]
\includegraphics[width=1\linewidth]{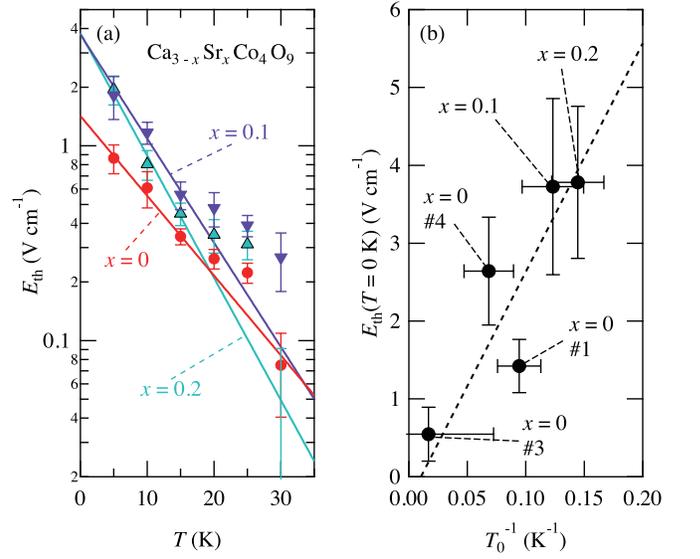}
\caption{(Color online).
(a) Temperature dependence of the threshold field of 
Ca$_{3-x}$Sr$_x$Co$_{4}$O$_9$ ($x=0, 0.1,0.2$)
shown in linear-log scale.
The solid lines represent the fitting results.
(b) Relation between  $E_{th}(0)$ and $T_0^{-1}$
for several samples of Ca$_{3-x}$Sr$_x$Co$_{4}$O$_9$.
The dashed line represents a linear $T_0^{-1}$ dependence of $E_{th}(0)$.}
\end{figure}

Figure 7(b) 
displays $E_{th}(0)$ as a function of $T_0^{-1}$
obtained for several samples of Ca$_{3-x}$Sr$_x$Co$_{4}$O$_9$,
roughly
showing a positive correlation between 
$E_{th}(0)$ and $T_0^{-1}$ in this cobaltite.
Here, 
$E_{th}(0)$ is sensitive to the amount of impurity, 
as is theoretically given by
$E_{th}(0) \propto c$  for the 
strong pinning limit,
where $c$ is the impurity concentration \cite{Maki1986}, 
and
$T_0^{-1} \propto \xi^{-1}=\xi^{-1}_{0}+\xi^{-1}_{\rm imp}$,
where $\xi_0 \simeq \hbar v_F/\pi k_BT_{\rm SDW}$
and $\xi_{\rm imp}$ is a mean free math 
determined by the amount of impurities \cite{Gruner}.
Since $\xi^{-1}_{\rm imp}$
represents an impurity concentration per unit length,
a relation of $E_{th}(0) \propto \xi^{-1}_{\rm imp} \propto 
T_0^{-1} + A$ ($A$ being a constant) is expected, 
and 
the experimental data roughly show this linear $T_0^{-1}$ dependence of $E_{th}(0)$
as depicted by the dashed line in fig. 7(b).
To further examine the impurity effect,
we  compare two characteristic lengths, $\xi_{0}$ and $\xi_{\rm imp}$.
Using the carrier concentration $n \simeq 6\times 10^{21}$ cm$^{-3}$ \cite{Schrade2015} 
and 
the effective mass $m^*\simeq 13m_0$ \cite{Wang2010}, 
$\xi_0$ is calculated as $\xi_0 \simeq 4$~nm.
On the other hand, 
$\xi_{\rm imp}$ is estimated as $\xi_{\rm imp} \simeq 4$~nm
for $x=0.1$ sample \cite{xiimp}, 
comparable to $\xi_0$,
indicating that the impurities certainly affect 
the temperature dependence of the threshold field in this system.
To examine more precisely, 
the impurity effect using both nonmagnetic and magnetic ions 
should be clarified as a future study.

The present study including the Sr substitution effect
certainly 
shows an intimate relationship between the low-temperature 
transport and the SDW order.
Interestingly, the  resistivity upturn reminiscent of the present system 
has also been observed in other misfit cobaltites
with different block-layer structures \cite{Wei2017}.
Thus nonlinear conduction should also be examined in such systems
as a future study
to clarify the ubiquitous nature among the layered cobalt oxides.
Moreover, since the low-temperature conductivity in Ca$_3$Co$_{4}$O$_9$
is suggested to be scaled to the Seebeck coefficient \cite{Hsieh2014},
the electric field effect on other transport properties 
including the Seebeck coefficient and the thermal conductivity
is also crucial to elucidate the inherent conduction mechanism 
far from equilibrium \cite{Stokes1984,Kriza1990}.

\section{summary}

We have studied the nonlinear conduction phenomena in 
the layered cobalt oxide Ca$_3$Co$_{4}$O$_9$ and 
the Sr-substituted systems.
Below the SDW transition temperature,
the electrical conductivity largely varies with both
the applied current and the  time,
indicating that the observed nonlinear conduction is 
caused by a dynamics of SDW.
We analyze the time dependence of the conductivity 
in terms of the low-field deformation and the 
high-field sliding motion of SDW,
and find that 
the temperature and Sr-content variations of the threshold fields 
are qualitatively understood within a
thermal fluctuation effect on the SDW.

\section*{acknowledgements}

We thank K. Kanai, Y. Maeno, S. Yonezawa for discussion and 
R. Adachi, H. Koike for experimental supports.
This work was supported by 
a Grant-in-Aid for 
Challenging Exploratory Research (No. 26610099)
from JSPS
 and a start-up research funding for young scientists from Tokyo University of Science.


\begin{thebibliography}{99}
\bibitem{xiimp}
The unit cell of the rocksalt-type Ca$_2$CoO$_3$ layer 
contains four Ca ions and 
the lattice constant 
is $a\sim b_1\sim0.5$~nm, leading to a rough estimation of the Sr-Sr distance of 
$0.5\times 30/4\simeq 4$ nm for $x=0.1$ sample.
\end{thebibliography}
\end{document}